\documentclass[twocolumn,aps,prl,superscriptaddress]{revtex4}
\usepackage{amsmath}
\usepackage{amssymb}
\usepackage{etoolbox}
\usepackage{graphicx}
\usepackage{color}

\newcommand{\be}{\begin{equation}}
\newcommand{\ee}{\end{equation}}
\def\ba{\begin{aligned}}
\def\ea{\end{aligned}}

\newcommand{\bea}{\begin{eqnarray}}
\newcommand{\eea}{\end{eqnarray}}

\newcommand{\titleinfo}{Anderson Localization on the Bethe Lattice: Nonergodicity of Extended States} 

\begin{document}

\title{\titleinfo}
\author{Andrea De Luca}
\email{andrea.deluca@lpt.ens.fr}
\affiliation{Laboratoire de Physique Th\'eorique de l'ENS and Institut de Physique
Theorique Philippe Meyer\\
 24, rue Lhomond 75005 Paris, France.}

\author{B. L. Altshuler}
\affiliation{Physics Department, Columbia University, 538 West 120th Street, New York, NY 10027, USA}
 
\author{V. E.  Kravtsov}
\affiliation{Abdus Salam International Center
for Theoretical Physics, Strada Costiera 11, 34151 Trieste, Italy}
\affiliation{L.D.Landau  Institute for Theoretical Physics, 2 Kosygina street, 119334 Moscow, Russia}

\author{A. Scardicchio}
\affiliation{Physics Department, Columbia University, 538 West 120th Street, New York, NY 10027, USA}
\affiliation{Abdus Salam International Center
for Theoretical Physics, Strada Costiera 11, 34151 Trieste, Italy}
\affiliation{Physics Department, Princeton University, Princeton, NJ 08544, USA}
\affiliation{INFN, Sezione di Trieste, Strada Costiera 11, 34151 Trieste, Italy. }

\begin{abstract}
Statistical analysis of the eigenfunctions of the Anderson tight-binding model with on-site disorder on regular random graphs strongly suggests that the extended states are multifractal at any finite disorder. The spectrum of fractal dimensions $f(\alpha)$ defined in Eq.(3), remains positive for $\alpha$ noticeably far from 1 even when the disorder is several times weaker than the one which leads to the Anderson localization, i.e.\ the ergodicity can be reached only in the absence of disorder. The one-particle multifractality on the Bethe lattice signals on a possible inapplicability of the equipartition law to a generic many-body quantum system as long as it remains isolated.
\end{abstract}


\maketitle

{\it Introduction.--- } Anderson localization (AL) \cite{50Anderson,Anderson58}, in its broad sense, is one of the central paradigms of quantum theory. Diffusion, which is a generic asymptotic behavior of classical random walks \cite{Einstein}, is inhibited in quantum case and under certain conditions it ceases to exist \cite{Anderson58}. This concerns quantum transport of noninteracting particles subject to quenched disorder as well as transport and relaxation in many-body systems. In the latter case the {\it many-body localization} (MBL) \cite{MBL} can be thought of as localization in the Fock space of Slater determinants, which play the role of lattice sites in a disordered  tight-binding model. In contrast to a  $d$-dimensional lattice, the structure of Fock space is hierarchical\cite{AGKL}: a two-body interaction couples a one-particle excitation with three one-particle excitations, which in turn are coupled with five-particle excitations, etc.
This structure resembles a random regular graph (RRG) - a finite size Bethe lattice (BL) without boundary. Interest to the problem of single particle AL on the BL \cite{Abou-Chacra, Abou-Chacra1} has recently revived \cite{Aiz-S, Aiz-War, Bir-Sem, Biroli, monthus2011anderson} largely in connection with MBL.
	It is a good approximation to consider  hierarchical lattices as trees
where any pair of sites is connected by only one path and loops are absent. Accordingly the sites in resonance with each other are much sparser than in ordinary $d>1$-dimensional lattices. As a result even the extended wave functions can occupy zero fraction of the BL, i.e.\ be {\it nonergodic}.
	The nonergodic extended states on 3D lattices where loops are abundant are commonly believed \cite{Weg, AKL,  KrMut, Mir-rev} to exist but only at the critical point of the AL transition.
\begin{figure}
\center{\includegraphics[width=0.7\linewidth]{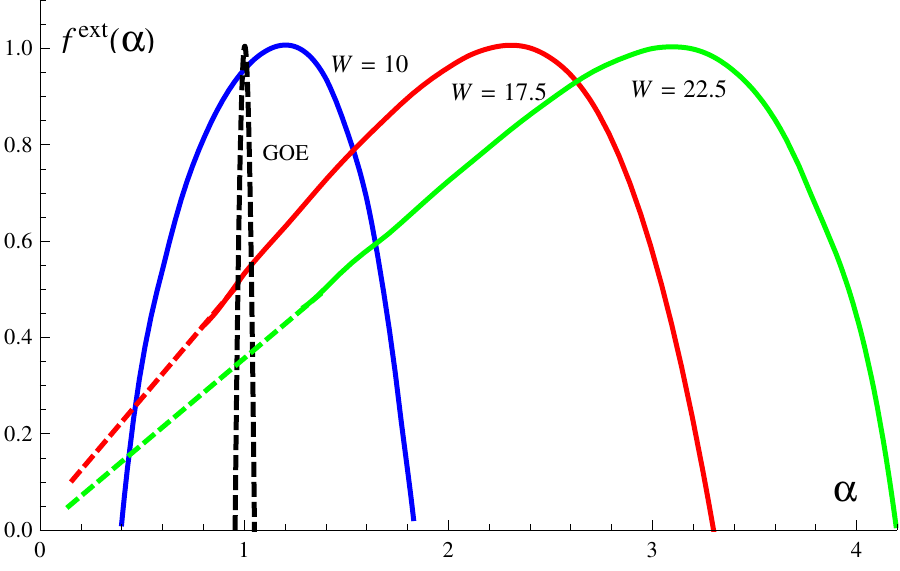}}
\caption{(Color online)
 Numerical results for $f(\alpha)$ on the RRG with the connectivity $K+1=3$ after linear extrapolation $f(\alpha,N)=f^{{\rm ext}}(\alpha)+c/\ln N$ to $1/\ln N\rightarrow 0$ for different values of disorder $W$. The dashed straight lines show the slope $k <1/2$ for the localized ($W=22.5$) and $k=1/2$  for the critical ($W=17.5$) states.} \label{Fig:f-num}
\end{figure}

In this paper we analyze the eigenstates of the Anderson model on RRG with connectivity $K+1$ ($K$ is commonly used to refer to the \emph{branching} of the corresponding BL) and $N$ sites:
\begin{equation}\label{AM}
H=-\sum_{<ij>}(c^\dag_i c_j+\mbox{h.c.})+\sum_i \varepsilon_i c^\dag_i c_i,
\end{equation}
where $\varepsilon_i\in[-W/2,W/2]$. A normalized wave function $\psi(i)$  ($i=1,...,N$) can be characterized by the moments $I_{q}=\sum_{i}|\psi(i)|^{2q}\propto N^{-\tau(q)}$~ \cite{Weg}  ($I_{1}=1$ for the normalization). 
One can define the ergodicity as the convergence in the limit $N\rightarrow\infty$ of the real space averaged
$|\psi(i)|^{2q}$ (equal to $I_{q}/N$) to its ensemble average value $\langle |\psi(i)|^{2q}\rangle=\langle I_{q}\rangle/N$.
This happens when
the fluctuations of $|\psi(i)|^{2}$ are relatively weak and $\langle |\psi(i)|^{2q} \rangle=a(q)\,\langle |\psi(i)|^{2}\rangle^{q}$ with $a(q)=O(N^{0})$. Since
$\psi(i)$ is normalized $\langle |\psi(i)|^{2}\rangle=N^{-1}$ and thus $I_{q}=N\langle |\psi(i)|^{2q} \rangle = a(q)\,N^{1-q}$, i.e. $\tau(q)=q-1$. The
latter condition turns out to be both necessary and sufficient for the convergence of $I_{q}$ to $\langle I_{q}\rangle$
(see Supplementary Materials for the discussion). Deviations of $\tau(q)$ from $q-1$ are signatures of the nonergodic state. If the ratio $D_{q}=\tau(q)/(q-1)$ depends on $q$, the wave function $\psi(i)$ is called {\it multifractal}. It is  customary to characterize $\psi(i)$ by the spectrum of fractal dimensions (SFD) $f(\alpha)$ related to $\tau(q)$ by the Legendre transform: $\tau(q)=q\alpha-f(\alpha)$ with $\alpha(q)$ being a solution to $f'(\alpha)=q$ (see supplemental material). Such a relationship follows from the definition of $f(\alpha)$, Eq.(3), and the saddle-point approximation in evaluating of the moments $I_{q}$ at large $\ln N$.

In this Letter we develop a method of extracting SFD $f(\alpha)$ from the numerical diagonalization of the Hamiltonian Eq.(\ref{AM}) on the RRG with finite number of sizes $N$ and  branching $K=2$. The multifractality is overshadowed by the fast oscillations $\phi_{{\rm osc}}(i)$ of $\psi(i)$  which should be separated from the smooth envelope $\psi_{{\rm en}}(i)$:
\be\label{prod-anz}
\psi(i)=\psi_{{\rm en}}(i)\,\phi_{{\rm osc}}(i).
\ee
Below we describe how to separate the statistics of $\psi_{{\rm en}}(i)$ and demonstrate that {\it at all strengths} $W$ of the on-site disorder the distribution function (DF) of $x=N\,|\psi_{{\rm en}}(i)|^{2}$ is consistent with the {\it multifractal ansatz} \cite{AKL, Mir-rev}, i.e.\ it can be expressed through SFD $f(\alpha)$ as:
\be\label{mult-anz}
P(x)=\frac{A}{x}\; N^{f(\alpha)-1},\;\;\;\alpha(x)=1-\ln x/\ln N,
\ee
where $A\sim O(N^{0})$ is the normalization constant. The SFD $f(\alpha)$ in Eq.(\ref{mult-anz}) is known \cite{Weg, Mir-rev} to be a convex function equal to 1 at its maximum, $f_{{\rm max}}=f(\alpha_{0})=1$. For ergodic states $f(\alpha) =-\infty$ unless $\alpha=1$ where $f(1)=1$, while a finite support $0<\alpha_{{\rm min}}<\alpha<\alpha_{{\rm max}}$ where $f(\alpha)>0$ is a signature of multifractality (nonergodicity).

We found that with decreasing disorder
$f(\alpha)$ evolves from almost triangular shape in the insulator to a steep parabolic shape concentrated near $\alpha=1$ (see Fig.\ref{Fig:f-num}).

Fractal behaviour of quantum dynamics on the disordered BL have been discussed previously. Transmission from the root to a given surface point of the Cayley tree turns out to be multifractal \cite{monthus2011anderson}. This surface multifractality of the extended states does not necessarily mean that bulk of the BL is multifractal: it is known, e.g., that in 2D the bulk multifractality is much weaker than the surface one \cite{Mir-rev}.
Our analysis of the results of exact numerical diagonalization of the Hamiltonian (1) on the RRG demonstrated that the extended wave functions are multifractal even in the bulk of the BL.
Hopefully the tools developed in \cite{Aiz-War,Aiz-S,aizenman2011resonant} might lead to a proof of the multifractality in the whole delocalized region, a possibility that we plan to explore in future work.

Authors of Ref.\cite{Biroli} analyzed numerically the statistics of the spectra and by population dynamics the distribution of the Green functions of the model (1) and conjectured a transition between extended-ergodic and extended nonergodic phases in addition to the Anderson transition. Contrarily, we do not see any evidence of the second transition and believe that the entire extended phase is nonergodic.

{\it Numerics on the BL: rectification and extrapolation.--- }
With the exception of the deeply localized states discussed below, analytical methods to address the problem of wave function statistics on BL are yet to be developed. One can try to access these statistics numerically by diagonalization of the Anderson model Eq.(\ref{AM}) on a RRG.
The first challenge along this route is the necessity to extract the statistics of the smooth envelope $\psi_{{\rm en}}(i)$ of the wave function $\psi(i)$, Eq.(\ref{prod-anz}). The short-range oscillations of $\phi_{{\rm osc}}(i)$ have nothing to do with AL but dominate the numerically obtained DF of $|\psi(i)|^{2}$ at small $|\psi(i)|^{2}$. This tail of the DF thus reflects the density of the nodes of $\phi_{{\rm osc}}(i)$ rather than the probability for $|\psi_{{\rm en}}(i)|^{2}$ to be small.

Since the scales of spatial dependencies of $\psi_{{\rm en}}(i)$ and $\phi_{{\rm osc}}(i)$ are so different, it is natural to assume that these two functions are statistically independent and $|\phi_{{\rm osc}}|^2$ is characterized by the Porter-Thomas DF of the Gaussian Orthogonal Ensemble (GOE) \cite{mehta} $P_{{\rm GOE}}(|\phi_{{\rm osc}}|^2)=(\sqrt{2\pi} |\phi_{{\rm osc}}|)^{-1}\exp\left(-|\phi_{{\rm osc}}|^2/2\right)$. Under these assumptions $\tilde\alpha=1-\ln(N|\psi|^2)/\ln N$, which is evaluated numerically, is a sum of two statistically independent random variables: $\tilde\alpha=\alpha(x)+\alpha_{{\rm osc}}$ where $\alpha(x)$ is given by Eq.\ (3) and $\alpha_{{\rm osc}}=-\ln|\phi_{{\rm osc}}|^2/\ln N$. The DF of $\tilde\alpha$ is thus a convolution of the DF $p(\alpha)=P(x(\alpha))x(\alpha)\ln N$ of $\alpha(x)$ with $\tilde P_{{\rm GOE}}(\alpha_{{\rm osc}})=\ln N\,(2\pi\,N^{\alpha_{{\rm osc}}})^{-\frac{1}{2}}\,\exp(N^{-\alpha_{{\rm osc}}}/2)$, i.e.\ $P(x)$ determined by (3) can be obtained (``rectified") from the DF of $\tilde\alpha$ and by the Laplace transform method (see supplemental material for details).

Another, even  bigger challenge is that Eq.(\ref{mult-anz}) is expected to hold only in the limit of a sufficiently large graph, i.e. $f(\alpha)$ can be determined only from the limit of $f(\alpha,N)=f(\alpha)+\delta_{N}f(\alpha)$ at $\ln N\rightarrow\infty$, where:
\be\label{f-alpha-N}
f(\alpha,N)=2-\alpha + \ln P(x)/\ln N,\;\;\;\; x=N^{1-\alpha}.
\ee

The biggest number of sites accessible to us was $N=32,000$. Increasing further $N$ does not buy much, as the computation time increases as $N^{3}$ while the finite-size correction $\delta_{N}f(\alpha)$ is small only as $1/\ln N$. However, we found a bright side of the slowness of the convergence: in the broad interval of $\alpha$ the correction $\delta_{N}f(\alpha)$ turns out to be linear in $1/\ln N$ to a surprisingly  high accuracy. This allowed us to make a  reliable {\it linear extrapolation} in $1/\ln N$ well beyond the numerical data.

\begin{figure}
\center{\includegraphics[width=0.8\linewidth]{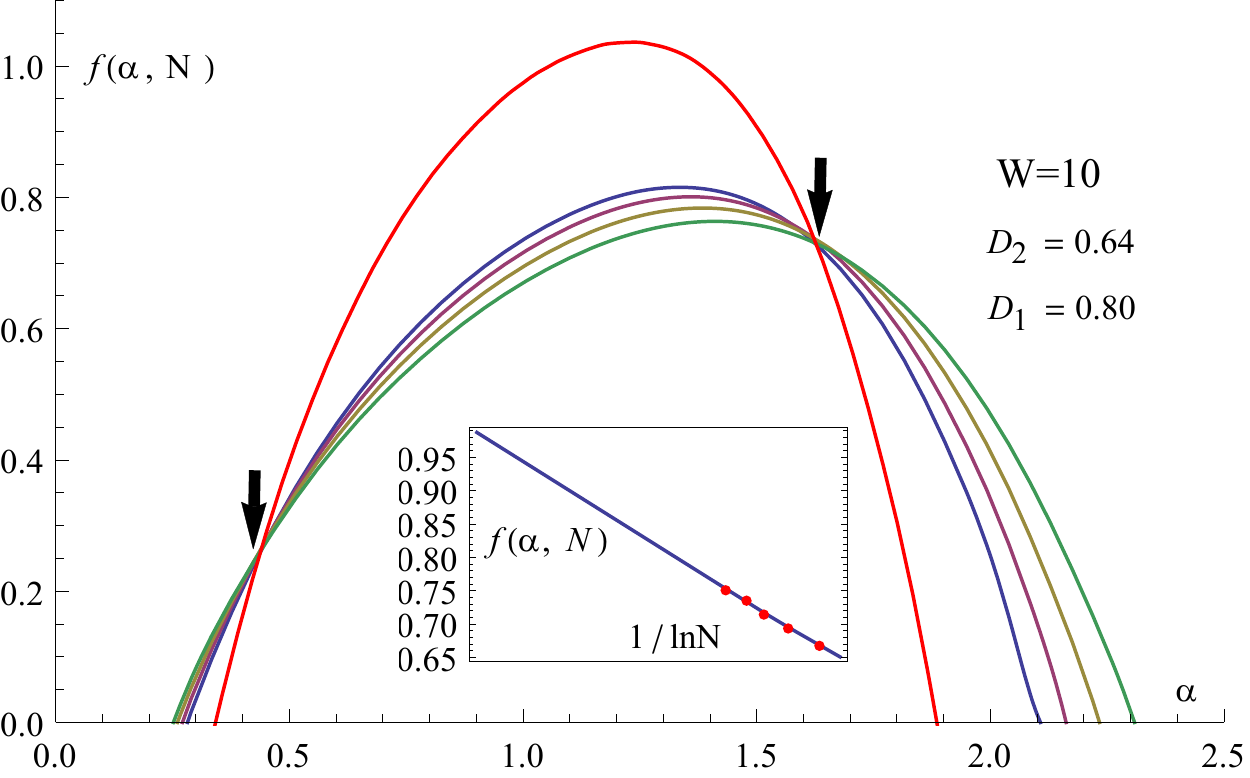}}
\caption{(Color online)
$N$-dependence of $f(\alpha,N)$ on the RRG with $K=2$ for $N=2,4,8,16\times 10^3$ (from green to blue in ascending order) in the extended phase $W=10$. The $f^{{\rm ext}}(\alpha)$ obtained by linear extrapolation of $f(\alpha,N)$ to $1/\ln N \rightarrow 0$ is shown by a thin  red line. The maximal value $f_{{\rm max}}\approx 1.03$ of $f^{{\rm ext}}(\alpha)$   is very close to the theoretical expectation $f_{{\rm max}}=1$.  We also show the fractal dimensions $D_{2}$ and $D_{1}=\lim_{q\rightarrow1+}D_{q}$ corresponding to $f^{{\rm ext}}(\alpha)$. The plots for different $N$ show  apparent fixed points at $\alpha\approx 0.5$ and $\alpha\approx 1.6$ indicated by arrows. Similar fixed points with $W$-dependent positions are seen at any strength of disorder studied.  This rules out that $f(\alpha,N)$ approaches at $\ln N\rightarrow\infty$ the ergodic limit $f(\alpha)=1$ at $\alpha=1$, $f(\alpha)=-\infty$ otherwise. In the insert: the linear extrapolation of $f(\alpha,N)$  for $\alpha=1.0$; the red points show $f(\alpha=1.0,N)$ at $N=2,4,8,16,32\times 10^{3}$.  } \label{Fig:fixed-points}
\end{figure}

\begin{figure}
\center{\includegraphics[width=0.8\linewidth]{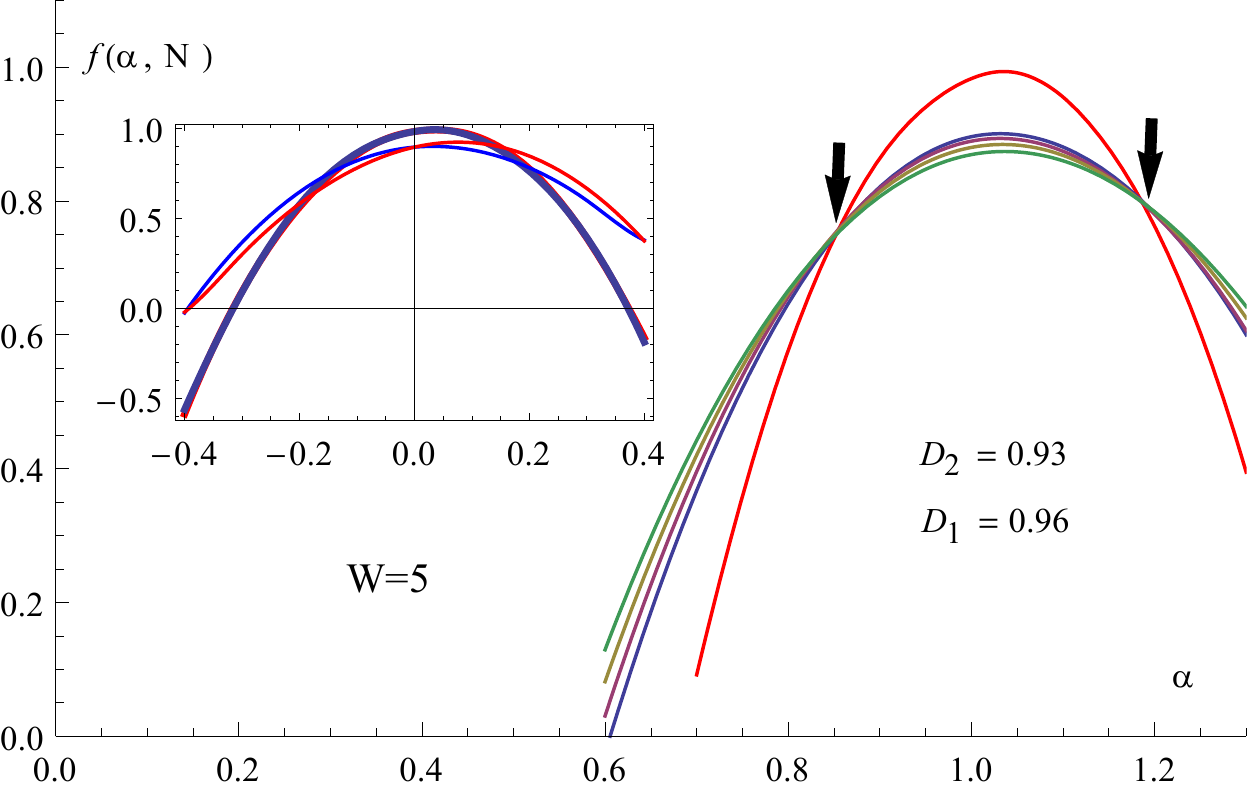}}
\caption{(Color online)
 $f^{{\rm ext}}(\alpha)$ obtained by linear extrapolation (see Fig.\ref{Fig:fixed-points})) of $f(\alpha,N)$ to $1/\ln N\rightarrow 0$ (red) and $f(\alpha,N)$ for $N=2,4,8,16\times 10^{3}$ at disorder strength $W=5$. The fixed points are shown by arrows. In the insert: verification of the symmetry Eq.(\ref{MF-sym}) for the extrapolated $f^{{\rm ext}}(\alpha)$ (coinciding blue and red thick curves), and for $f(\alpha,N)$ at $N=16\times 10^{3}$ (distinctly different thin blue and red curves).} \label{Fig:MF5}
\end{figure}

{\it Fixed points in the $N$-dependence of $f(\alpha,N)$---. }
We numerically diagonalized the Hamiltonian Eq.\ (\ref{AM}) for the regular graph with connectivity $K+1=3$ (which does not contain boundary sites) extracting about $2\%$ of the states around the center $E=0$ of the spectrum, and evaluated
$f(\alpha,N)$ for $N=32,16,8,4,2\times 10^{3}$ at several disorder strengths $W$, both above the Anderson transition at $W_{c}=17.5$ and below it, down to $W=5$.
Plots of $f(\alpha)$ for $W=10$ and $W=5$ are shown in Fig.\ref{Fig:fixed-points} and Fig.\ref{Fig:MF5}.

An important observation is the existence of two fixed points $\alpha_{+}$ and $\alpha_{-}$ (shown by arrows): $f(\alpha_{+},N)$ and $f(\alpha_{-},N)$ are essentially $N-independent$. This rules out the possibility of $f(\alpha)$ evolving with further increase of $N$ into  a sharp parabola at $\alpha=1$ (dashed line in Fig.\ref{Fig:f-num}).
In addition to that we verified to a high degree of accuracy that $\delta_{N}f(\alpha)$ is linear in $1/\ln N$, at least for $\alpha_{-}<\alpha<\alpha_{+}$. The insert of Fig.\ref{Fig:fixed-points} demonstrates that the values of $f(\alpha=1,N)$ being plotted as a function of $1/\ln N$ for $N=32,16,8,4,2\times 10^{3}$ (red points) form an almost ideal straight line which can be prolonged down to $1/\ln N =0$. This is how we obtained the extrapolated SFD $f^{{\rm ext}}(\alpha)$. It was already mentioned that the maximal value of $f(\alpha)$ in Eq.(\ref{mult-anz}) should be 1.  It is not the case for $f(\alpha,N)$ as one can see from Figs.\ref{Fig:fixed-points},\ref{Fig:MF5}. After extrapolation, however, the maxima of $f^{{\rm ext}}(\alpha)$ turn out to be much closer to 1: $f^{{\rm ext}}_{{\rm max}}=0.99,\; 1.03,\; 1.01,\; 1.00$
for $W=5,\;10,\;17.5,\;22.5$, respectively. We just conclude that the extrapolation passed an important and non-trivial test for consistence.

{\it Verification of the symmetry of $f(\alpha)$.--- } Another important observation providing us with the additional confidence in the validity of the extrapolation  is the symmetry
of SFD and DF:
\bea\label{MF-sym}
f(1+\alpha)&=&f(1-\alpha)+\alpha,\\ P(x)&=&x^{-3}\,P(x^{-1}) \label{dual}.
\eea
One can use Eq.(\ref{mult-anz}) to check that Eqs.(\ref{MF-sym}),(\ref{dual}) follow from each other.
Log-normal distribution found for weakly multifractal states in 2D disordered systems \cite{EfetFal} is one of the examples of this symmetry. A relation similar to Eq.(\ref{dual})  was proven   for the DF of the local density of states $\rho(i,\varepsilon)$ in a one-dimensional chain \cite{AlPrig} and a variety of systems (e.g.\ short and long disordered wires, 2D and 3D disordered systems) described by the nonlinear sigma-model \cite{Mir-Fyod-sym, Mir-Fyod-sym1}.
The precise conditions of validity of Eq.(\ref{dual}) for the \emph{individual eigenfunctions} are yet to be formulated. It does not hold for the localized eigenstates, while for weakly multifractal extended states in 2D systems it is valid \cite{EfetFal}. A vast numerical evidence of the validity of Eq.(\ref{MF-sym}) for the multifractal states at the Anderson transition point in 2D and 3D systems was reported \cite{Mir-rev}. In the insert of Fig.\ref{Fig:MF5} we present the separate plots of $f^{{\rm ext}}(1+\alpha)$ and $f^{{\rm ext}}(1-\alpha)+\alpha$  for the weakest disorder we studied, $W=5$ (deep in the region of extended states, the fractal dimensions $D_{1}$ and $D_{2}$ are very close to 1). One can see that the two curves are indistinguishable in the interval $-0.4<\alpha<0.4$, while $f(1+\alpha,N=16\times 10^{3})$ and $f(1-\alpha,N=16\times 10^{3})+\alpha$ differ noticeably.

In the localized regime $W>W_{c}=17.5$ and at the critical point $W=W_{c}$ the shape of the SFD is approximately triangular (see Fig.\ref{Fig:f-num}):
\be\label{PL}
f^{{\rm ext}}(\alpha)=k\,\alpha\,\theta(1-k\,\alpha),\;\;\;\theta(z)=
\begin{cases}
1 & \mbox{if } z>0\\
0 & \mbox{if } z<0.
\end{cases}
\ee
The slope $k$ depends on the disorder $k=k(W)$ with $k(W_{c})$ is very close to $1/2$. Note, that the only linear $f(\alpha)$ allowed by Eq.(\ref{MF-sym}) is the one with $k=1/2$. Thus one concludes that the critical states with not very small amplitude of wave function (not very large $\alpha$) for $W=W_c$ obey Eq.(\ref{MF-sym}). At the same time, the most abundant critical states around the maximum value of $f(\alpha)$ reached at $\alpha=\alpha_{0}\approx 2.3$ clearly violate the symmetry Eq.(\ref{MF-sym}). Indeed $f(1-\alpha)$ is defined only for $\alpha<1$ and thus according to Eq.(\ref{MF-sym}) $f(1+\alpha)$ should make no sense for $\alpha>1$. In the localized regime we found that $\alpha_{0}$ increases with disorder, while $k(W)\approx \alpha_{0}^{-1}$ decreases below $1/2$. This is in a clear contradiction with  Eq.(\ref{MF-sym}).

{\it Power-law DF for strongly localized states on BL.--- } Our numeric-based conclusions on the regime of strong localization $W\gg W_{c}$ fully agree with the analytical results which follow from the locator expansion \cite{Abou-Chacra} (also see supplemental material). In the {\it shortest path approximation} \cite{MedinaKardar,Derrida} one obtains: $\psi^{(0)}(i)=\prod_{j\in p_{0,i}}(\varepsilon_{0}-\varepsilon_{i})^{-1}$, where $\psi^{(0)}(i)$ is the eigenfunction of the Hamiltonian Eq.(\ref{AM}) on BL which at $W=\infty$ is located on the site 0, and $p_{0,i}$ is the shortest path connecting sites 0 and $i$. One can show that within this approximation the DF $P(x)$ can be represented as:
\be\label{P-x-I}
P(x)=I(1,0)-N^{1-\kappa}\,I(1-m,\kappa).
\ee
Here $m=\ln N/\ln K$ is the BL radius, $\kappa=\ln(W/2)/\ln K$, and
\be\label{I}
I(p,q)=\frac{1}{\sqrt{Nx^{3}}}\int_{B}\frac{ds}{4\pi i}\,\frac{s^{p}\,(x\,N^{2q-1})^{\frac{s}{2}}}{s-K^{\kappa(s-1)+1}}.
\ee
The contour $B\in (r-i\infty, r+i\infty)$ is parallel to the imaginary axis and crosses the real axis at $s_{-}<r < s_{+}$, where $s_{\pm}$ are the larger and the smaller of the only two real roots of the equation
\be\label{pole-eq}
s=K^{\kappa(s-1)+1}.
\ee
These roots can be shown to exist as long as $\kappa>\kappa_{c}=\ln(\kappa_{c}\ln K)+\ln(eK)/\ln K$, which can be rewritten as
\be\label{AT-point}
W\geq W_{c}=2e\,K\,\ln(W_{c}/2)\approx 2 e\,K\,\ln(e\,K).
\ee
Solution of Eq.(\ref{AT-point}) is nothing but the critical disorder of Ref. \cite{Anderson58}  (see Eq.(84) there, see also the ``upper limit critical condition'' of Ref.\cite{Abou-Chacra, Abou-Chacra1}).
Since $x=N|\psi(i)|^{2}<N$ in the first term of R.H.S. of Eq.(\ref{P-x-I}) one can deform the contour of integration in Eq.(\ref{I}) to encircle the pole $s=s_{+}$. For the second term, $N^{1-\kappa}\,I(1-m,\kappa)$, this can be done only provided that $x(N/K)^{2\kappa}<N$. Under this condition the two terms cancel each other and $P(x)\equiv 0$. In the opposite case $x>N(N/K)^{-2\kappa}$, the integral $I(1-m,\kappa)$ in Eq.(\ref{I}) is determined by the poles $s=s_{-}$ and $s=0$. Within the region of validity of the shortest path approximation $\frac{s_{-}}{2}\,\ln x \sim \frac{2\kappa}{W}\,\ln N\ll 1$ the two contributions cancel each other, i.e. $P(x)=I(1,0)$. Finally within the shortest path approximation, for $W\gg W_{c}$ we have:
\be\label{P-sh-path}
P(x)=\frac{\theta(x-x_{{\rm min}})}{N^{2}}\,\left(\frac{x}{N}\right)^{(s_{+}-3)/2},
\ee
where
$x_{{\rm min}}=N^{1-2\kappa}\,K^{2\kappa} \Rightarrow \alpha_{{\rm max}}=2\kappa\,(1-m^{-1})\approx 2 \kappa.$
%
Using the definition of $\alpha(x)$ Eq.(\ref{mult-anz}) one obtains from the power-law DF Eq.(\ref{P-sh-path}) the linear SFD $f(\alpha)$, Eq.(\ref{PL}) with:
\be\label{f-alpha-lin}
k(W)=\frac{1}{2}(1-s_{+}),
\ee
truncated at $\alpha>\alpha_{{\rm max}}$. Note that for $\kappa\gg 1$ (i.e. for $W\gg 2K$), Eq.(\ref{pole-eq}) yields $s_{+}\approx 1-\kappa^{-1}$, so that the condition $k(W)\alpha_{{\rm max}}=1$ encoded in Eq.(\ref{PL}) is satisfied at large disorder ($\kappa\gg 1$) and large system size ($m\gg 1$). Power-law distributions of wave function coefficients have been observed in many-body systems also in the delocalized region (see \cite{DeLuca} where a criterion for ergodicity breaking based on them was proposed).

{\it Conclusion.--- } We developed an effective method for extracting statistics of the smooth envelopes $\psi_{en}(i)$ of random eigenfunctions  of the Anderson model (1) on RRG and to extrapolate these results from RRG to the BL with an infinite number of sites. Our results strongly suggests that DF of  $|\psi_{en}(i)|^2$ in the limit $N\to\infty$  indeed converges to the form Eq.(3) regardless to the strength of disorder. As long as the states are localized the spectrum of fractal dimensions turns out to be triangular:  $f(0)=0$ and  the linear $f(\alpha)$ is well described by Eq.(7). The slope $k$ increases as the disorder $W$ decreases and reaches its maximal possible value $k_c=1/2$ at the Anderson transition point $W=W_c=17.5$. With further decrease of the disorder below the critical one $f(\alpha)$ gradually crosses over to the parabolic shape typical for weak multifractality: the two roots of $f(\alpha)$ become positive $0<\alpha_{min}<1<\alpha_{max}$ ($\alpha_{min}\to 0^+$ as $W\to W_c^-$). However even for $W$ several times smaller than $W_c$ both $\alpha_{min}$ and $\alpha_{max}$ turn out to be quite far from 1, while the ergodicity would imply that $\alpha_{min},\alpha_{max}\to1$. We conclude that the nonergodicity and multifractality persist in the entire region of delocalized states $0<W<W_c$, and the only critical point is the point of the Anderson localization transition.

It goes without saying that only RRG with not too big $N$ are accessible for the numerical analysis and one has to deal with $f(\alpha,N)$ determined by Eq.(4). However the existing data allow us to exclude the possibility that the observed nonergodicity is a finite size effect. Our confidence is based, among other things, on the existence of two fixed points:  $f(\alpha,N)$ is $N$-independent at $\alpha=\alpha_-(W)<1$ and $\alpha=\alpha_+(W)>1$. The extrapolation of $f(\alpha,N)$ to $N\to\infty$ in the interval $\alpha_-\leq\alpha\leq\alpha_+$  turned out to be tremendously reliable. It is thus hard to imagine how $f(\alpha,N)$ could evolve to the ergodic limit with further increase of $N$.

Another argument in favour of the true nonergodicity is that the behaviour of $f(\alpha,N)$ Eq.\ (\ref{f-alpha-N}) is \emph{not a critical behavior}: $f(\alpha,N)$ depends on both $N$ and $W$ in a broad range of these variables. Indeed the critical behaviour, which was analytically predicted in \cite{MF1991,MF1997} for \emph{sparse random matrices} (SRM) implies that at $W<W_c$ the eigenvectors are ergodic or multifratcal correspondingly for $N>N_c(W)$ and for $N<N_c(W)$ (the critical volume $N_c(W)$ diverges as $W\to W_c$). Therefore $f(\alpha,N)$ depends either on $N$ (in the critical regime) or on $W$, but never on both $N$ and $W$. The reasons why the results of refs.\ \cite{MF1991,MF1997} do not apply to the RRG eigenvectors will be discussed elsewhere.

The absence of ergodicity for the dynamics of the one-particle Anderson model on the BL, in light of the possible connection with the many-body dynamics, suggests serious implications on the statistical mechanics of isolated systems with a large number of degrees of freedom. If the same phenomenon occurs in the many-body case, the equipartition law is likely not to be valid exactly even for strongly non-integrable systems.

{\it Acknowledgments.} We would like to thank Yan V.\ Fyodorov, Michael Aizenman, Eugene Bogomolny and  Markus Mueller for useful discussions and Giulio Biroli for discussions on his work \cite{Biroli} which was one of the main motivations for us to undertake the work presented in this paper. AS is grateful to the Graduate College and Initiative for Theoretical Sciences of the City University of New York for financial support; BLA  acknowledges financial support from Triangle de la Physique through the project DISQUANT.

\bibliography{albethe.bib}

\begin{thebibliography}{27}%
\makeatletter
\providecommand \@ifxundefined [1]{%
 \@ifx{#1\undefined}
}%
\providecommand \@ifnum [1]{%
 \ifnum #1\expandafter \@firstoftwo
 \else \expandafter \@secondoftwo
 \fi
}%
\providecommand \@ifx [1]{%
 \ifx #1\expandafter \@firstoftwo
 \else \expandafter \@secondoftwo
 \fi
}%
\providecommand \natexlab [1]{#1}%
\providecommand \enquote  [1]{``#1''}%
\providecommand \bibnamefont  [1]{#1}%
\providecommand \bibfnamefont [1]{#1}%
\providecommand \citenamefont [1]{#1}%
\providecommand \href@noop [0]{\@secondoftwo}%
\providecommand \href [0]{\begingroup \@sanitize@url \@href}%
\providecommand \@href[1]{\@@startlink{#1}\@@href}%
\providecommand \@@href[1]{\endgroup#1\@@endlink}%
\providecommand \@sanitize@url [0]{\catcode `\\12\catcode `\$12\catcode
  `\&12\catcode `\#12\catcode `\^12\catcode `\_12\catcode `\%12\relax}%
\providecommand \@@startlink[1]{}%
\providecommand \@@endlink[0]{}%
\providecommand \url  [0]{\begingroup\@sanitize@url \@url }%
\providecommand \@url [1]{\endgroup\@href {#1}{\urlprefix }}%
\providecommand \urlprefix  [0]{URL }%
\providecommand \Eprint [0]{\href }%
\providecommand \doibase [0]{http://dx.doi.org/}%
\providecommand \selectlanguage [0]{\@gobble}%
\providecommand \bibinfo  [0]{\@secondoftwo}%
\providecommand \bibfield  [0]{\@secondoftwo}%
\providecommand \translation [1]{[#1]}%
\providecommand \BibitemOpen [0]{}%
\providecommand \bibitemStop [0]{}%
\providecommand \bibitemNoStop [0]{.\EOS\space}%
\providecommand \EOS [0]{\spacefactor3000\relax}%
\providecommand \BibitemShut  [1]{\csname bibitem#1\endcsname}%
\let\auto@bib@innerbib\@empty
\bibitem [{\citenamefont {Abrahams}(2010)}]{50Anderson}%
  \BibitemOpen
  \bibfield  {author} {\bibinfo {author} {\bibfnamefont {E.}~\bibnamefont
  {Abrahams}},\ }\href@noop {} {\emph {\bibinfo {title} {50 years of Anderson
  localization}}},\ Vol.~\bibinfo {volume} {24}\ (\bibinfo  {publisher} {World
  Scientific},\ \bibinfo {year} {2010})\BibitemShut {NoStop}%
\bibitem [{\citenamefont {Anderson}(1958)}]{Anderson58}%
  \BibitemOpen
  \bibfield  {author} {\bibinfo {author} {\bibfnamefont {P.~W.}\ \bibnamefont
  {Anderson}},\ }\href@noop {} {\bibfield  {journal} {\bibinfo  {journal}
  {Phys. Rev.}\ }\textbf {\bibinfo {volume} {109}},\ \bibinfo {pages} {1492}
  (\bibinfo {year} {1958})}\BibitemShut {NoStop}%
\bibitem [{\citenamefont {Einstein}(1905)}]{Einstein}%
  \BibitemOpen
  \bibfield  {author} {\bibinfo {author} {\bibfnamefont {A.}~\bibnamefont
  {Einstein}},\ }\href@noop {} {\bibfield  {journal} {\bibinfo  {journal}
  {Annalen der physik}\ }\textbf {\bibinfo {volume} {322}},\ \bibinfo {pages}
  {549} (\bibinfo {year} {1905})}\BibitemShut {NoStop}%
\bibitem [{\citenamefont {Basko}\ \emph {et~al.}(2006)\citenamefont {Basko},
  \citenamefont {Aleiner},\ and\ \citenamefont {Altshuler}}]{MBL}%
  \BibitemOpen
  \bibfield  {author} {\bibinfo {author} {\bibfnamefont {D.}~\bibnamefont
  {Basko}}, \bibinfo {author} {\bibfnamefont {I.}~\bibnamefont {Aleiner}}, \
  and\ \bibinfo {author} {\bibfnamefont {B.}~\bibnamefont {Altshuler}},\
  }\href@noop {} {\bibfield  {journal} {\bibinfo  {journal} {Ann. Phys.}\
  }\textbf {\bibinfo {volume} {321}},\ \bibinfo {pages} {1126} (\bibinfo {year}
  {2006})}\BibitemShut {NoStop}%
\bibitem [{\citenamefont {Altshuler}\ \emph {et~al.}(1997)\citenamefont
  {Altshuler}, \citenamefont {Gefen}, \citenamefont {Kamenev},\ and\
  \citenamefont {Levitov}}]{AGKL}%
  \BibitemOpen
  \bibfield  {author} {\bibinfo {author} {\bibfnamefont {B.~L.}\ \bibnamefont
  {Altshuler}}, \bibinfo {author} {\bibfnamefont {Y.}~\bibnamefont {Gefen}},
  \bibinfo {author} {\bibfnamefont {A.}~\bibnamefont {Kamenev}}, \ and\
  \bibinfo {author} {\bibfnamefont {L.~S.}\ \bibnamefont {Levitov}},\
  }\href@noop {} {\bibfield  {journal} {\bibinfo  {journal} {Phys. Rev. Lett.}\
  }\textbf {\bibinfo {volume} {78}},\ \bibinfo {pages} {2803} (\bibinfo {year}
  {1997})}\BibitemShut {NoStop}%
\bibitem [{\citenamefont {Abou-Chacra}\ \emph {et~al.}(1973)\citenamefont
  {Abou-Chacra}, \citenamefont {Thouless},\ and\ \citenamefont
  {Anderson}}]{Abou-Chacra}%
  \BibitemOpen
  \bibfield  {author} {\bibinfo {author} {\bibfnamefont {R.}~\bibnamefont
  {Abou-Chacra}}, \bibinfo {author} {\bibfnamefont {D.}~\bibnamefont
  {Thouless}}, \ and\ \bibinfo {author} {\bibfnamefont {P.}~\bibnamefont
  {Anderson}},\ }\href@noop {} {\bibfield  {journal} {\bibinfo  {journal} {J.
  Phys. C: Solid State Phys.}\ }\textbf {\bibinfo {volume} {6}},\ \bibinfo
  {pages} {1734} (\bibinfo {year} {1973})}\BibitemShut {NoStop}%
\bibitem [{\citenamefont {Abou-Chacra}\ and\ \citenamefont
  {Thouless}(1974)}]{Abou-Chacra1}%
  \BibitemOpen
  \bibfield  {author} {\bibinfo {author} {\bibfnamefont {R.}~\bibnamefont
  {Abou-Chacra}}\ and\ \bibinfo {author} {\bibfnamefont {D.}~\bibnamefont
  {Thouless}},\ }\href@noop {} {\bibfield  {journal} {\bibinfo  {journal} {J.
  Phys. C: Solid State Phys.}\ }\textbf {\bibinfo {volume} {7}},\ \bibinfo
  {pages} {65} (\bibinfo {year} {1974})}\BibitemShut {NoStop}%
\bibitem [{\citenamefont {Aizenman}\ \emph {et~al.}(2006)\citenamefont
  {Aizenman}, \citenamefont {Sims},\ and\ \citenamefont {Warzel}}]{Aiz-S}%
  \BibitemOpen
  \bibfield  {author} {\bibinfo {author} {\bibfnamefont {M.}~\bibnamefont
  {Aizenman}}, \bibinfo {author} {\bibfnamefont {R.}~\bibnamefont {Sims}}, \
  and\ \bibinfo {author} {\bibfnamefont {S.}~\bibnamefont {Warzel}},\
  }\href@noop {} {\bibfield  {journal} {\bibinfo  {journal} {Commun. Math.
  Phys.}\ }\textbf {\bibinfo {volume} {264}},\ \bibinfo {pages} {371} (\bibinfo
  {year} {2006})}\BibitemShut {NoStop}%
\bibitem [{\citenamefont {Aizenman}\ and\ \citenamefont
  {Warzel}(2011)}]{Aiz-War}%
  \BibitemOpen
  \bibfield  {author} {\bibinfo {author} {\bibfnamefont {M.}~\bibnamefont
  {Aizenman}}\ and\ \bibinfo {author} {\bibfnamefont {S.}~\bibnamefont
  {Warzel}},\ }\href@noop {} {\bibfield  {journal} {\bibinfo  {journal}
  {Europhys. Lett.}\ }\textbf {\bibinfo {volume} {96}},\ \bibinfo {pages}
  {37004} (\bibinfo {year} {2011})}\BibitemShut {NoStop}%
\bibitem [{\citenamefont {Biroli}\ \emph {et~al.}(2010)\citenamefont {Biroli},
  \citenamefont {Semerjian},\ and\ \citenamefont {Tarzia}}]{Bir-Sem}%
  \BibitemOpen
  \bibfield  {author} {\bibinfo {author} {\bibfnamefont {G.}~\bibnamefont
  {Biroli}}, \bibinfo {author} {\bibfnamefont {G.}~\bibnamefont {Semerjian}}, \
  and\ \bibinfo {author} {\bibfnamefont {M.}~\bibnamefont {Tarzia}},\
  }\href@noop {} {\bibfield  {journal} {\bibinfo  {journal} {Prog. Theor. Phys.
  Suppl.}\ }\textbf {\bibinfo {volume} {184}},\ \bibinfo {pages} {187}
  (\bibinfo {year} {2010})}\BibitemShut {NoStop}%
\bibitem [{\citenamefont {Biroli}\ \emph {et~al.}(2012)\citenamefont {Biroli},
  \citenamefont {Ribeiro-Teixeira},\ and\ \citenamefont {Tarzia}}]{Biroli}%
  \BibitemOpen
  \bibfield  {author} {\bibinfo {author} {\bibfnamefont {G.}~\bibnamefont
  {Biroli}}, \bibinfo {author} {\bibfnamefont {A.}~\bibnamefont
  {Ribeiro-Teixeira}}, \ and\ \bibinfo {author} {\bibfnamefont
  {M.}~\bibnamefont {Tarzia}},\ }\href@noop {} {\bibfield  {journal} {\bibinfo
  {journal} {arXiv preprint arXiv:1211.7334}\ } (\bibinfo {year}
  {2012})}\BibitemShut {NoStop}%
\bibitem [{\citenamefont {Monthus}\ and\ \citenamefont
  {Garel}(2011)}]{monthus2011anderson}%
  \BibitemOpen
  \bibfield  {author} {\bibinfo {author} {\bibfnamefont {C.}~\bibnamefont
  {Monthus}}\ and\ \bibinfo {author} {\bibfnamefont {T.}~\bibnamefont
  {Garel}},\ }\href@noop {} {\bibfield  {journal} {\bibinfo  {journal} {J.
  Phys. A: Math. Theor.}\ }\textbf {\bibinfo {volume} {44}},\ \bibinfo {pages}
  {145001} (\bibinfo {year} {2011})}\BibitemShut {NoStop}%
\bibitem [{\citenamefont {Wegner}(1981)}]{Weg}%
  \BibitemOpen
  \bibfield  {author} {\bibinfo {author} {\bibfnamefont {F.}~\bibnamefont
  {Wegner}},\ }\href@noop {} {\bibfield  {journal} {\bibinfo  {journal} {Eur.
  Phys. J. B}\ }\textbf {\bibinfo {volume} {44}},\ \bibinfo {pages} {9}
  (\bibinfo {year} {1981})}\BibitemShut {NoStop}%
\bibitem [{\citenamefont {Altshuler}\ \emph {et~al.}(1986)\citenamefont
  {Altshuler}, \citenamefont {Kravtsov},\ and\ \citenamefont {Lerner}}]{AKL}%
  \BibitemOpen
  \bibfield  {author} {\bibinfo {author} {\bibfnamefont {B.}~\bibnamefont
  {Altshuler}}, \bibinfo {author} {\bibfnamefont {V.}~\bibnamefont {Kravtsov}},
  \ and\ \bibinfo {author} {\bibfnamefont {I.}~\bibnamefont {Lerner}},\
  }\href@noop {} {\bibfield  {journal} {\bibinfo  {journal} {JETP Lett.}\
  }\textbf {\bibinfo {volume} {43}} (\bibinfo {year} {1986})}\BibitemShut
  {NoStop}%
\bibitem [{\citenamefont {Kravtsov}\ \emph {et~al.}(1994)\citenamefont
  {Kravtsov}, \citenamefont {Lerner}, \citenamefont {Altshuler},\ and\
  \citenamefont {Aronov}}]{KrMut}%
  \BibitemOpen
  \bibfield  {author} {\bibinfo {author} {\bibfnamefont {V.}~\bibnamefont
  {Kravtsov}}, \bibinfo {author} {\bibfnamefont {I.}~\bibnamefont {Lerner}},
  \bibinfo {author} {\bibfnamefont {B.}~\bibnamefont {Altshuler}}, \ and\
  \bibinfo {author} {\bibfnamefont {A.}~\bibnamefont {Aronov}},\ }\href@noop {}
  {\bibfield  {journal} {\bibinfo  {journal} {Phys. Rev. Lett.}\ }\textbf
  {\bibinfo {volume} {72}},\ \bibinfo {pages} {888} (\bibinfo {year}
  {1994})}\BibitemShut {NoStop}%
\bibitem [{\citenamefont {Evers}\ and\ \citenamefont {Mirlin}(2008)}]{Mir-rev}%
  \BibitemOpen
  \bibfield  {author} {\bibinfo {author} {\bibfnamefont {F.}~\bibnamefont
  {Evers}}\ and\ \bibinfo {author} {\bibfnamefont {A.~D.}\ \bibnamefont
  {Mirlin}},\ }\href@noop {} {\bibfield  {journal} {\bibinfo  {journal} {Rev.
  Mod. Phys.}\ }\textbf {\bibinfo {volume} {80}},\ \bibinfo {pages} {1355}
  (\bibinfo {year} {2008})}\BibitemShut {NoStop}%
\bibitem [{\citenamefont {Aizenman}\ and\ \citenamefont
  {Warzel}(2013)}]{aizenman2011resonant}%
  \BibitemOpen
  \bibfield  {author} {\bibinfo {author} {\bibfnamefont {M.}~\bibnamefont
  {Aizenman}}\ and\ \bibinfo {author} {\bibfnamefont {S.}~\bibnamefont
  {Warzel}},\ }\href@noop {} {\bibfield  {journal} {\bibinfo  {journal} {J.
  Eur. Math. Soc.}\ }\textbf {\bibinfo {volume} {15}},\ \bibinfo {pages} {1167}
  (\bibinfo {year} {2013})}\BibitemShut {NoStop}%
\bibitem [{\citenamefont {Mehta}(2004)}]{mehta}%
  \BibitemOpen
  \bibfield  {author} {\bibinfo {author} {\bibfnamefont {M.~L.}\ \bibnamefont
  {Mehta}},\ }\href@noop {} {\emph {\bibinfo {title} {Random matrices}}},\
  Vol.\ \bibinfo {volume} {142}\ (\bibinfo  {publisher} {Academic press},\
  \bibinfo {year} {2004})\BibitemShut {NoStop}%
\bibitem [{\citenamefont {Fal'ko}\ and\ \citenamefont
  {Efetov}(1995)}]{EfetFal}%
  \BibitemOpen
  \bibfield  {author} {\bibinfo {author} {\bibfnamefont {V.~I.}\ \bibnamefont
  {Fal'ko}}\ and\ \bibinfo {author} {\bibfnamefont {K.~B.}\ \bibnamefont
  {Efetov}},\ }\href@noop {} {\bibfield  {journal} {\bibinfo  {journal} {Phys.
  Rev. B}\ }\textbf {\bibinfo {volume} {52}},\ \bibinfo {pages} {17413}
  (\bibinfo {year} {1995})}\BibitemShut {NoStop}%
\bibitem [{\citenamefont {Altshuler}\ and\ \citenamefont
  {Prigodin}(1989)}]{AlPrig}%
  \BibitemOpen
  \bibfield  {author} {\bibinfo {author} {\bibfnamefont {B.}~\bibnamefont
  {Altshuler}}\ and\ \bibinfo {author} {\bibfnamefont {V.}~\bibnamefont
  {Prigodin}},\ }\href@noop {} {\bibfield  {journal} {\bibinfo  {journal} {Zh.
  Eksp. Teor. Fiz.}\ }\textbf {\bibinfo {volume} {95}},\ \bibinfo {pages} {348}
  (\bibinfo {year} {1989})}\BibitemShut {NoStop}%
\bibitem [{\citenamefont {Mirlin}\ and\ \citenamefont
  {Fyodorov}(1994{\natexlab{a}})}]{Mir-Fyod-sym}%
  \BibitemOpen
  \bibfield  {author} {\bibinfo {author} {\bibfnamefont {A.~D.}\ \bibnamefont
  {Mirlin}}\ and\ \bibinfo {author} {\bibfnamefont {Y.~V.}\ \bibnamefont
  {Fyodorov}},\ }\href@noop {} {\bibfield  {journal} {\bibinfo  {journal}
  {Phys. Rev. Lett.}\ }\textbf {\bibinfo {volume} {72}},\ \bibinfo {pages}
  {526} (\bibinfo {year} {1994}{\natexlab{a}})}\BibitemShut {NoStop}%
\bibitem [{\citenamefont {Mirlin}\ and\ \citenamefont
  {Fyodorov}(1994{\natexlab{b}})}]{Mir-Fyod-sym1}%
  \BibitemOpen
  \bibfield  {author} {\bibinfo {author} {\bibfnamefont {A.~D.}\ \bibnamefont
  {Mirlin}}\ and\ \bibinfo {author} {\bibfnamefont {Y.~V.}\ \bibnamefont
  {Fyodorov}},\ }\href@noop {} {\bibfield  {journal} {\bibinfo  {journal} {J.
  Phys. I}\ }\textbf {\bibinfo {volume} {4}},\ \bibinfo {pages} {655} (\bibinfo
  {year} {1994}{\natexlab{b}})}\BibitemShut {NoStop}%
\bibitem [{\citenamefont {Medina}\ and\ \citenamefont
  {Kardar}(1992)}]{MedinaKardar}%
  \BibitemOpen
  \bibfield  {author} {\bibinfo {author} {\bibfnamefont {E.}~\bibnamefont
  {Medina}}\ and\ \bibinfo {author} {\bibfnamefont {M.}~\bibnamefont
  {Kardar}},\ }\href@noop {} {\bibfield  {journal} {\bibinfo  {journal} {Phys.
  Rev. B}\ }\textbf {\bibinfo {volume} {46}},\ \bibinfo {pages} {9984}
  (\bibinfo {year} {1992})}\BibitemShut {NoStop}%
\bibitem [{\citenamefont {Miller}\ and\ \citenamefont
  {Derrida}(1994)}]{Derrida}%
  \BibitemOpen
  \bibfield  {author} {\bibinfo {author} {\bibfnamefont {J.~D.}\ \bibnamefont
  {Miller}}\ and\ \bibinfo {author} {\bibfnamefont {B.}~\bibnamefont
  {Derrida}},\ }\href@noop {} {\bibfield  {journal} {\bibinfo  {journal} {J.
  Stat. Phys.}\ }\textbf {\bibinfo {volume} {75}},\ \bibinfo {pages} {357}
  (\bibinfo {year} {1994})}\BibitemShut {NoStop}%
\bibitem [{\citenamefont {De~Luca}\ and\ \citenamefont
  {Scardicchio}(2013)}]{DeLuca}%
  \BibitemOpen
  \bibfield  {author} {\bibinfo {author} {\bibfnamefont {A.}~\bibnamefont
  {De~Luca}}\ and\ \bibinfo {author} {\bibfnamefont {A.}~\bibnamefont
  {Scardicchio}},\ }\href@noop {} {\bibfield  {journal} {\bibinfo  {journal}
  {Europhys. Lett.}\ }\textbf {\bibinfo {volume} {101}},\ \bibinfo {pages}
  {37003} (\bibinfo {year} {2013})}\BibitemShut {NoStop}%
\bibitem [{\citenamefont {Fyodorov}\ and\ \citenamefont
  {Mirlin}(1991)}]{MF1991}%
  \BibitemOpen
  \bibfield  {author} {\bibinfo {author} {\bibfnamefont {Y.~V.}\ \bibnamefont
  {Fyodorov}}\ and\ \bibinfo {author} {\bibfnamefont {A.~D.}\ \bibnamefont
  {Mirlin}},\ }\href@noop {} {\bibfield  {journal} {\bibinfo  {journal} {Phys.
  Rev. Lett.}\ }\textbf {\bibinfo {volume} {67}},\ \bibinfo {pages} {2049}
  (\bibinfo {year} {1991})}\BibitemShut {NoStop}%
\bibitem [{\citenamefont {Mirlin}\ and\ \citenamefont
  {Fyodorov}(1997)}]{MF1997}%
  \BibitemOpen
  \bibfield  {author} {\bibinfo {author} {\bibfnamefont {A.~D.}\ \bibnamefont
  {Mirlin}}\ and\ \bibinfo {author} {\bibfnamefont {Y.~V.}\ \bibnamefont
  {Fyodorov}},\ }\href@noop {} {\bibfield  {journal} {\bibinfo  {journal}
  {Phys. Rev. B}\ }\textbf {\bibinfo {volume} {56}},\ \bibinfo {pages} {13393}
  (\bibinfo {year} {1997})}\BibitemShut {NoStop}%
\end{thebibliography}%

\clearpage
\newpage

\clearpage 
\setcounter{equation}{0}%
\setcounter{figure}{0}%
\setcounter{table}{0}%
\renewcommand{\thetable}{S\arabic{table}}
\renewcommand{\theequation}{S\arabic{equation}}
\renewcommand{\thefigure}{S\arabic{figure}}

\onecolumngrid

\begin{center}
{\Large Supplementary Material for EPAPS \\ 
\titleinfo
}
\end{center}

\section{Ergodicity condition}
It is natural to define the ergodicity of random eigenstates $\psi(i)$ as the vanishing in the limit $N\rightarrow\infty$   difference between the real space average of $|\psi(i)|^{2q}$   (equal to $I_{q}/N$ ) and its ensemble average $\langle |\psi(i)|^{2q}\rangle=\langle I_{q}\rangle/N$.
The difference between the two mean values can be characterized by the normalized ensemble averaged square of this difference $\eta_{q}$:
\be\label{eta}
\eta_{q}=\frac{\left\langle \left[ N^{-1}I_{q}-\langle |\psi(i)|^{2q}\rangle\right]^{2}\right\rangle}{\langle |\psi(i)|^{2q}\rangle^{2}}=
\frac{\left\langle I_{q}^{2}\right\rangle-\langle I_{q}\rangle^{2}}{\langle I_{q} \rangle^{2}}.
\ee
By normalization of eigenstates $\eta_{1}=0$. The non-trivial test for ergodicity is the value of $\eta_{q}$ at $q>1$.

The mean square moment of the participation ratio $\langle I_{q}^{2}\rangle$  can be written as
\be \label{mean-square}
\langle I_{q}^{2}\rangle = \sum_{i,j}\left\langle|\psi(i)|^{2q}\,|\psi(j)|^{2q} \right\rangle = N\sum_{i}\left\langle|\psi(i)|^{2q}\,|\psi(0)|^{2q} \right\rangle.
\ee
The correlation function $\left\langle|\psi(i)|^{2q}\,|\psi(0)|^{2q} \right\rangle$  is determined by the linear distance $r_{i}$ of the site $i$ from the site $0$ or by the number $N_{i}=K^{r_{i}}$  of the sites, which distance from the site 0 does not exceed $r_{i}$ . Denoting
\be\label{korr}
\langle |\psi(i)|^{2q}\,|\psi(0)|^{2q} \rangle = \langle |\psi(0)|^{2q}\rangle^{2}\;F_{q}\left(\frac{N}{N_{i}}\right),
\ee
and using Eqs.(\ref{eta})-(\ref{korr}) we present $\eta_{q}$ as:
\be\label{devia}
\eta_{q}=\frac{1}{N}\sum_{i=1}^{N}\left[F_{q}\left(\frac{N}{N_{i}}\right)-1\right].
\ee
It is safe to assume that if $N_{i}=O(1)$ then the correlation function Eq.(\ref{korr}) obeys the {\it fusion rule}:
\be\label{fusion}
\langle |\psi(i)|^{2q}\,|\psi(0)|^{2q} \rangle \approx \langle |\psi(0)|^{4q}\rangle =N^{-1}\,\langle I_{2q} \rangle \sim N^{-\tau(2q)-1},
\ee
while for $N_{i}=O(N)$ the correlation between $|\psi(i)|^{2q}$ and $|\psi(0)|^{2q}$ is negligible and one should apply the {\it decomposition rule}:
\be\label{decomp}
\langle |\psi(i)|^{2q}\,|\psi(0)|^{2q} \rangle \approx  \langle |\psi(0)|^{2q} \rangle^{2}=N^{-2}\langle I_{q}\rangle^{2}\sim N^{-2\tau(q)-2}.
\ee
In the multifractal regime it is natural to assume for an arbitrary $N_{i}$  a power-like interpolation between Eq.(\ref{fusion}) and Eq.\ref{decomp}):
\be\label{power-interpol}
\langle |\psi(i)|^{2q}\,|\psi(0)|^{2q} \rangle \sim N^{-a(q)}\,\left( \frac{N}{N_{i}}\right)^{-b(q)}.
\ee
The exponents $a(q)$  and $b(q)$ can be determined from Eqs.(\ref{fusion}),(\ref{decomp}):
\be \label{a-b}
a(q)+b(q)=1+\tau(2q);\;\;\;\;a(q)=2+2\tau(q).
\ee
Therefore $b(q)=\tau(2q)-2\tau(q)-1$,  and the function $F_{q}(u)$ in Eq.(\ref{korr}) can be written as:
\be\label{FF}
F_{q}(u)=C_{q}\, u^{1-\tau(2q)+2\tau(q)}
\ee
with some pre-factor $C_{q}=O(N^{0})$.

Note that, if $\tau(q)=q-1$ then $\tau(2q)-2\tau(q)=1$ and $F_{q}(u)={\rm const.}$ . According to Eq.(\ref{korr}) this means that the correlation between  $|\psi(i)|^{2q}$ and $|\psi(0)|^{2q}$  can be neglected for {\it all} $N_{i}>O(1)$ , i.e. $C_{q}=1$ and only a few terms in the sum Eq.(\ref{devia}) over $N_{i}$   contribute to $\eta_{q}$. Therefore for $\tau(q)q-1$, $\eta_{q}=O(N^{-1})$ and the state is ergodic.

\begin{figure}
\center{\includegraphics[width=0.6\linewidth]{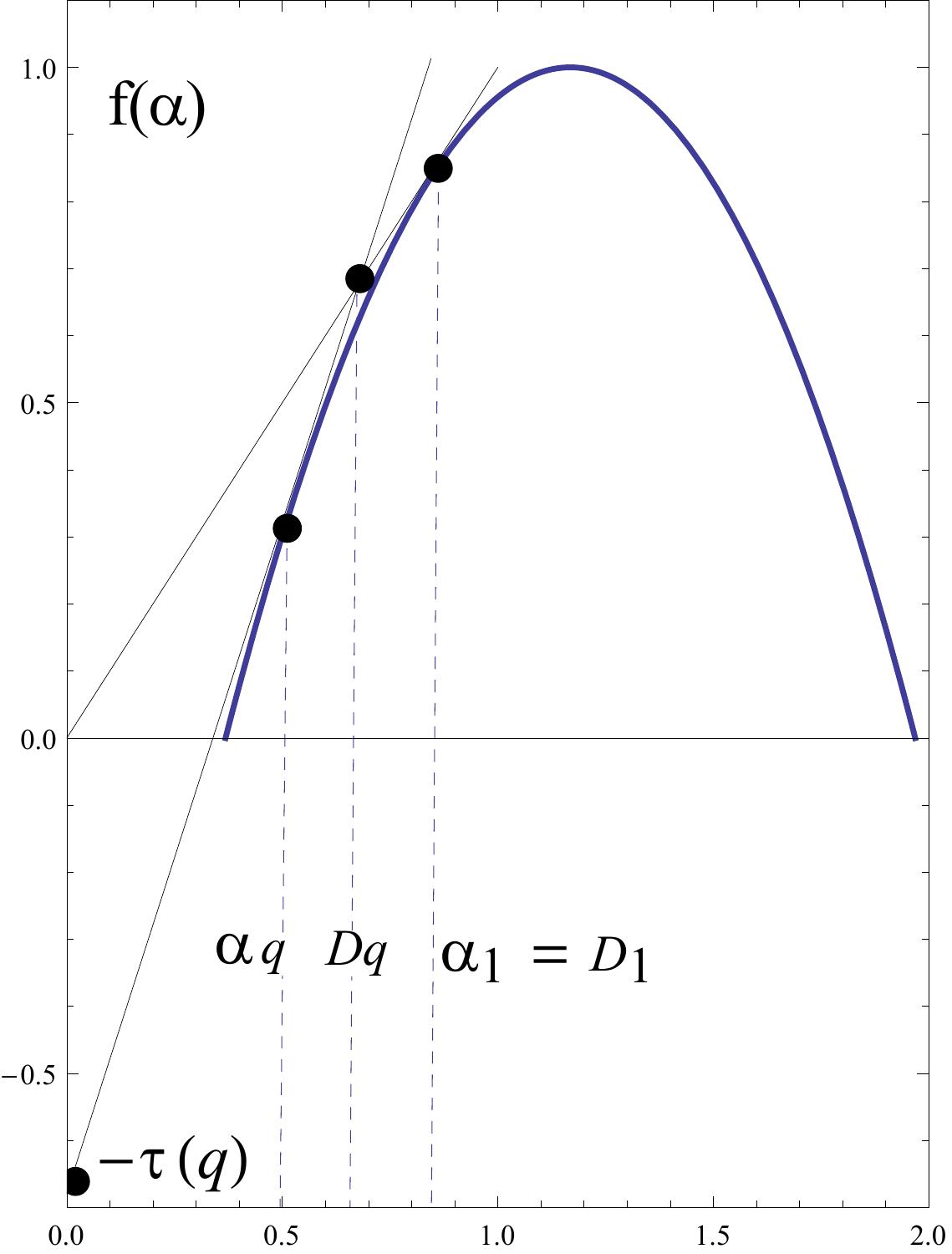}}
\caption{(Color online)
The function $f(\alpha)$ (thick blue solid line) and the Legendre transform to obtain $\tau(q)$: the black thin solid lines $y_{q}(\alpha)=q(\alpha-\alpha_{q})+f(\alpha_{q})=q\alpha-\tau(q)$ and $y_{1}(\alpha)=\alpha$ are tangential to the function $f(\alpha)$  at $\alpha=\alpha_{q}$ and $\alpha=\alpha_{1}$; the intersection of $y_{q}(\alpha)$ with the $y$-axis is equal to $-\tau(q)$. The fractal dimension $D_{q}=\tau(q)/(q-1)$ is given by $\alpha$ at the intersection of the lines $y_{q}(\alpha)$ and $y_{1}(\alpha)$. For a convex $f(\alpha)$ and $q>1$ one obtains $\alpha_{1}>D_{q}>\alpha_{q}$. The derivative $D'_{q}=dD_{q}/dq $ satisfies the relation $D'_{q}\,(q-1)=d\tau(q)/dq-D_{q}=\alpha_{q}-D_{q}$, and is thus negative for all $q>1$.   } \label{Fig:ineq}
\end{figure}

What happens if $\tau(q)\neq q-1$? Figure \ref{Fig:ineq} demonstrates that the convexity of the function $f(\alpha)$   in Eq.(\ref{mult-anz}) implies that $0\leq D_{2q}\leq D_{q}\leq 1$. Using this inequality it is easy to show that the combination $\tau(2q)-2\tau(q)=D_{2q}(2q-1)-2D_{q}(q-1)$ can take values only between 0 and 1:
\be\label{ineq}
0\leq \tau(2q)-2\tau(q)\leq 1.
\ee
Then the sum in Eq.(\ref{devia}) is dominated by $N_{i}=O(N)$ and thus can be replaced by an integral and evaluated using Eq.(\ref{FF}):
\be\label{sum-integr}
\eta_{q}\approx \int_{O(N^{-1})}^{1}du\,\left[ F_{q}(u)-1\right]=\frac{C_{q}}{\tau(2q)-2\tau(q)}-1+O(N^{-1}).
\ee
As was already mentioned at $\tau(q)=q-1$ both the numerator and the denominator in the fraction in Eq.(\ref{sum-integr}) are equal to 1, and the ergodicity condition $\lim_{N\rightarrow \infty}\eta_{q}=0$ is fulfilled. For $q=1$, the normalization of eigenfunctions requires $I_{1}=1$, and $\eta_{1}=0$. According to Eq.(\ref{sum-integr}) this results in $C_{1}=\tau(2)=D_{2}$ in all the regimes.
For $\tau(q)\neq (q-1)$ and $q>1$ there is no reason for $C_{q}$ to be equal to $\tau(2q)-2\tau(q)$, hence $\eta_{q}=O(1)$ is non-zero, and the ergodicity is violated.

\section{Distribution function in the forward scattering approximation}
For large disorder, the wave function can be written employing the locator expansion, from which we get
\begin{equation}
 \label{locator}
 \psi(i) = \sum_p \prod_{j \in p} \frac{1}{\varepsilon_0 - \varepsilon_j}
\end{equation}
where the sum runs over the path from $0$ to $i$. This expression simplifies on the BL where there is only one shortest $p$ path connecting two points. Let $n$ be the length of the path $p$. It is convenient to pass to $x_n=N|\psi(i)|^2$ and study the distribution of $\ln x_n$, as this is a sum of i.i.d.\ random variables. For simplicity we assume $\varepsilon_0=0$ and defining dimensionless $y_j=((W/2)/\varepsilon_j)^2$:
\begin{equation}
\ln x_n=\ln N+n \ln ((2/W)^2)+\sum_{j=1}^n\ln y_j.
\end{equation}
For the case of box distribution $\varepsilon_i \in [-W/2, W/2]$, $y_j>1$ and $\xi_n=\ln x_n-n\ln (N^{1/n}(2/W)^2)=\sum_{j}\ln y_j>0$ (in the case of more general on-site disorder, one should resort to Laplace transform but this technicality does not change the calculations substantially). We find
\begin{equation}
\label{eq:supmatpy}
p(y)=\frac{1}{2y^{3/2}}\theta(y-1).
\end{equation}
The power law tail at large $y$ is a common feature of any distribution and arises from the small denominators, which inhibit the existence of the average of $y$.

As usual, the Laplace transform of the sum of i.i.d.\ variables is the $n$-th power of the Laplace transform of that of a single variable which in this case is:
\begin{eqnarray}
R(s)&=&\int_0^\infty d\ln y\ e^{-s\ln y} p(\ln y)=\nonumber\\
&=&\int_1^\infty dy\ y^{-s}\frac{1}{2y^{3/2}}=\frac{1}{1+2s}.
\end{eqnarray}
So by taking the $n$-th power and inverting the Laplace transform we have formally:
\begin{equation}
P_n(\xi_n)=\int_B\frac{ds}{2\pi i} e^{s\xi_n}R(s)^n,
\end{equation}
where the Bromwich path $B$ passes to the right of the only singularity of the integrand, ($s=-1/2$ in the case of the box distribution).

Therefore the distribution of the $x_n=N(2/W)^{2n}e^{\xi_n}$ is
\begin{equation}
P_n(x_n)=\frac{1}{x_n}\left. P_n(\xi_n)\right|_{\xi_n=\ln x_n-\ln N+2n\ln (W/2)}
\end{equation}
so
\begin{equation}
\label{eq:supmatPn}
P_n(x_n)=\frac{1}{x_n}\int_B\frac{ds}{2\pi i}\left(\frac{x_n}{N}\right)^s(W/2)^{2ns}R(s)^n.
\end{equation}
We find now the probability distribution $P(x)$ by summing over the events that the given observation site $i$ belongs to the $n$-th generation:
\begin{equation}
\label{eq:supmatP}
P(x)=\sum_{n=0}^{\ln N/\ln K}\frac{K^{n-1}(K+1)}{N}P_n(x).
\end{equation}
The sum over $n$ can be performed exactly and we get the result
\be \label{P-k-forward}
P(x)=\frac{1}{2N^{\frac{1}{2}}\,x^{\frac{3}{2}}}\int_{B}\frac{d s}{2\pi i}\,(x/N)^{\frac{s}{2}}\frac{1-[(W/2)^{(s-1)}\,K/s]^{m}}{1-(W/2)^{(s-1)}\,K/s},
\ee
where $m = \ln N/\ln K + 1$. By introducing $\kappa=\ln(W/2)/\ln K$ we can rewrite (\ref{P-k-forward}) in the form of Eq.s\ (7) and (8) of the main text. Notice that the integrand is singular only for $s = 0$, since the other poles of the denominators are canceled out by zeros of the numerators.
The contour $B\in (r-i\infty, r+i\infty)$ has to be parallel to the imaginary axis with $r > 0$. In order to simplify the analysis, we chooses $s_{-}<r < s_{+}$, where $s_{\pm}$ are the larger and the smaller real root of the equation
\be\label{pole-eq}
s=K\,(W/2)^{(s-1)}=K^{\kappa(s-1)+1},
\ee
which is Eq.(10) of the main text.

\section{Rectification of the distribution function}

As discussed in the text, the distributions $P(x)$ for which the multifractal ansatz holds, describe the smooth envelope $\psi_{en}$ of the fast oscillating wave function $\psi$. This is, for some values of $x$, very much different from the numerically obtained distribution function ${\cal P}(N|\psi^{2}|)$ of the values of the wave function. For example, according to the multifractal ansatz, there is always a minimal statistically relevant $|\psi_{en}|^{2}=N^{-\alpha_{{\rm max}}}$ while ${\cal P}(N|\psi^{2}|)$ does not have this feature as $\psi$ can be arbitrarily close to 0 for finite $N$, due to interference effects.

We explain here the method, alternative to the existing ones and better suited for the Bethe lattice, to recover $P_{en}(x_{en})$ and from this, $f(\alpha)$. The numerical estimation of the fractal spectrum encoded in the function $f(\alpha)$ is usually a complicated task due to the fluctuation of the eigenstates on the scale of the lattice length. This fact can be seen from the function $P(x)$ that always presents a square-root behavior $x^{-1/2}$ at small $x$.
Other approaches are known to overcome this difficulty usually based on a real-space renormalization procedure at large wavelengths, usually called \textit{box counting}.
In our case, this procedure clashes with the exponential growth of the BL so that even for the largest sizes we can numerically achieve the spatial extension of the system remains rather small (the diameter of our largest system counts about 16 nodes).
For this reason we follow a different method. It is based on the assumption that the variable $x_{ED} = N|\psi|^2$, coming from exact diagonalization, can be factorized into two independent random variables
\begin{equation}
 \label{xdec}
 x_{ED} = x_{en} x_{GOE}
\end{equation}
where $x_{GOE}$ corresponds to fast oscillations in the Gaussian Orthogonal Ensemble with the distribution function $P_{GOE}(x_{GOE})=e^{-x_{GOE}/2}/\sqrt{2\pi x_{GOE}}$. Switching to logarithmic variables $\ln x_{ED}=\ln x_{en}+\ln x_{GOE}$, $P_{en}(\ln x_{en})$ is computed inverting the convolution of the distribution functions ${\cal P}(\ln x_{ED})=P_{en}(\ln x_{en})*P_{GOE}(\ln x_{GOE})$.
The distribution of $\ln x_{ED}$ can be obtained numerically by binning the eigenvectors, while the distribution of $\ln x_{GOE}$ is explicitly known. In this way, the distribution of $\ln x_{en}$ can be derived efficiently with the help of fast-Laplace transform
\begin{equation}
 \label{fft}
 \mathcal{Q}_{en}(k) = \frac{\mathcal{Q}_{ED}(k)}{\mathcal{Q}_{GOE} (k)}=\frac{2^{-i k-a^{2}k^{2}}\Gamma\left(\frac{1}{2}\right)}{\Gamma\left(\frac{1}{2}+ i k\right)}\,\mathcal{Q}_{ED}(k)
\end{equation}
where $\mathcal{Q}(k)$ generically indicates the Laplace transforms of the distributions $P(\ln x)$.
The only difficulty comes from the fact that the data of the distribution of $\ln x$ are affected by errors which spoil the behavior of the Laplace transform at large $k$. The result is that the right-hand side of \eqref{fft} explodes at large $k$, making the inversion rather unstable.
To avoid this problem, we smoothed out the data of $\mathcal{P}_{ED}(\ln x)$ with a Gaussian kernel with a characteristic width $a$. This adds an additional Gaussian factor $e^{- a^2 k^2}$ in the right-hand side of  \eqref{fft} which ensures convergence. Ideally, the original equation is recovered only when the width of the Gaussian kernel $a$ is sent to zero. However, we checked that
the results are sufficiently robust when the width is decreased until the numerical errors become too relevant ($a^2 \gtrsim 0.1$).
\end{document}